\documentclass[prl,a4paper,twocolumn,showpacs,preprintnumbers,floatfix,english]{revtex4}
\usepackage[utf8]{inputenc}
\usepackage[T1]{fontenc}
\usepackage{babel}
\usepackage{fancyhdr}
\usepackage{ifpdf}
\usepackage{graphicx}
\usepackage{color}
\usepackage{lineno}
\usepackage{pslatex}

\usepackage{bm}
\usepackage{dcolumn}

\usepackage{amsmath}
\usepackage{amssymb}
\usepackage{amscd}

\ifpdf
\usepackage{epstopdf}
\usepackage[%
  pdftitle={lGe2Se3},%
  pdfauthor={},%
  pdfsubject={lGe2Se3},%
  pdfkeywords={lGe2Se3},%
  pdfstartview=FitH,%
  bookmarks=true,%
  bookmarksopen=true,%
  breaklinks=true,%
  colorlinks=true,%
  linkcolor=blue,anchorcolor=blue,%
  citecolor=blue,filecolor=blue,%
  menucolor=blue,pagecolor=blue,%
  urlcolor=blue]{hyperref}
\else
\usepackage[%
  pdftitle={lGe2Se3},%
  pdfauthor={},%
  pdfsubject={lGe2Se3},%
  pdfkeywords={lGe3Se3},%
  breaklinks=true,%
  colorlinks=true,%
  linkcolor=blue,anchorcolor=blue,%
  citecolor=blue,filecolor=blue,%
  menucolor=blue,pagecolor=blue,%
  urlcolor=blue]{hyperref}
\fi

\include{bibsymb}

\begin{document}
%\twocolumn[\hsize\textwidth\columnwidth\hsize\csname @twocolumnfalse\endcsname
\draft

%\title{Fragility and super-strong character of non-stoichiometric chalcogenides: implication on melt homogeneization}
\title{Reading the structure of amorphous materials from diffraction patterns and neighbor distribution functions}

\author{M. Micoulaut}
\affiliation{Sorbonne Université, Laboratoire de Physique Th\'{e}orique de la Mati\`ere Condensée, CNRS UMR 7600, 4 Place Jussieu, 75252 Paris Cedex 05, France}

\affiliation{}

\date{\today}

\begin{abstract}
An exact analytical expression for the static structure factor $S(k)$ in disordered materials is derived from Fourier transformed neighbor distribution decompositions in real space, and permits to reconstruct the function $S(k)$ in an iterative fashion. The result is successfully compared to experimental data of archetypal glasses or amorphous materials (GeS$_2$, As$_2$Se$_3$, GeTe), and links quantitatively knowledge of structural information on short and intermediate -range order with the motifs found on the diffraction patterns in reciprocal space. The approach furthermore reveals that only a limited number of neighbor shells is sufficient to reasonably describe the structure factor for $k>$2~\AA$^{-1}$. In the limit of the high momentum transfer, the oscillation characteristics of the interference function are related with new informations on the short-range order of disordered materials. 

\end{abstract}

\maketitle
Diffraction methods form the basis of our common understanding of materials. The determination of crystal structures has provided a foundation for considerable characterization of chemical bonding, cohesion, and functionalities in solid- or liquid-state science with applications in e.g. physico-chemistry, biology and geology. In contrast to crystals which display well-defined diffraction patterns and sharp Bragg peaks induced by the periodic arrangement of atoms, glasses and liquids exhibit broad and diffuse peaks that result from their intrinsic disordered structure. For such materials and with changing momentum transfer $k$, there have been many attempts to extract from the relevant resulting coherent diffraction function, that is, the static structure factor $S(k)$,  information on short- (SRO) and intermediate-range order (IRO) and a conventional means uses probabilistic atomic distribution functions to infer the microscopic structure \cite{review}. 
\par
Within this framework, the key quantity is the pair correlation function $g(r)$ which measures the probability of finding an atom at a position $r$ relative to a reference atom taken to be at the origin \cite{hansen}. Relating in detail the real space structure properties encoded in $g(r)$ and the shape and peak positions or amplitudes of the measured $S(k)$ remains a challenging task but has inspired a certain recent number of insightful contributions in the field \cite{contrib1,contrib2,contrib3,contrib4}, many years after the early attempts of Warren \cite{warren} and Bernal \cite{bernal}. In fact, only certain specific signatures can be detected and related to some structural features. The first sharp diffraction peak (FSDP) at low momentum transfer ($k \simeq$1~\AA$^{-1}$) reveals for instance some ordering at intermediate length scales \cite{elliott,bychkovfsdp} that might manifest in e.g. voids and depends on pressure and temperature. Similarly, principal peak (PP) characteristics are thought to be linked to the presence of extended range order \cite{Nature_salmon,salmon2}. Such features continue, however, to be discussed given that the established correlations remain essentially at a qualitative level and are largely material dependent.
\par
Here we show that with the assistance of computer simulated structural models of glasses, the decomposition into neighbor distribution functions in real space permits to reconstruct as a series expansion the structure factor, and an exact analytical expression for $S(k)$ is obtained from Fourier transformation. Such expressions exist for imperfect crystals with weak disorder \cite{guinier} but here it is the first time that a similar approach is used for disordered materials such as glasses. This not only links quantitatively information on ordering with characteristics of the structure factor $S(k)$, but also reveals what aspects of structure directly  influence tyical features observed experimentally at different momentum transfer. We first concentrate on the amorphous system As$_2$Se$_3$ before investigating the degree of generality of the findings on other chalcogenide glasses such as GeS$_2$ and GeTe (see Supplementary materials section \cite{suppl}). The high $k$ limit of the analytical expression leads to another exact result that permits to fully characterize the typical oscillations observed in the experimental interference function $I(k)$=$k$[$S(k)-1]$. 
\par
The starting point of the approach uses the definition \cite{hansen} of the structure factor given by :
\begin{eqnarray}
\label{sk}
S(k)-1=4\pi\rho_0\int_0^\infty r^2[g(r)-1]{\frac {\sin(kr)}{kr}}dr
\end{eqnarray}
where $\rho_0$ is the atomic number density. The system is considered as being isotropic and homogeneous so that integration can be applied on the vector norms $r=\vert {\bf r}\vert$ only, this being particularly adapted for disordered systems simulated in a cubic box of size $L$ down to wavelength somewhat larger than $2\pi/L$. It is well known however that equ. (\ref{sk}) fails to describe properly crystalline or lamellar phases, as recently reemphasized in a pedagogical fashion \cite{struct}. 
\par
\begin{figure}
\includegraphics*[width=\linewidth, keepaspectratio=true]{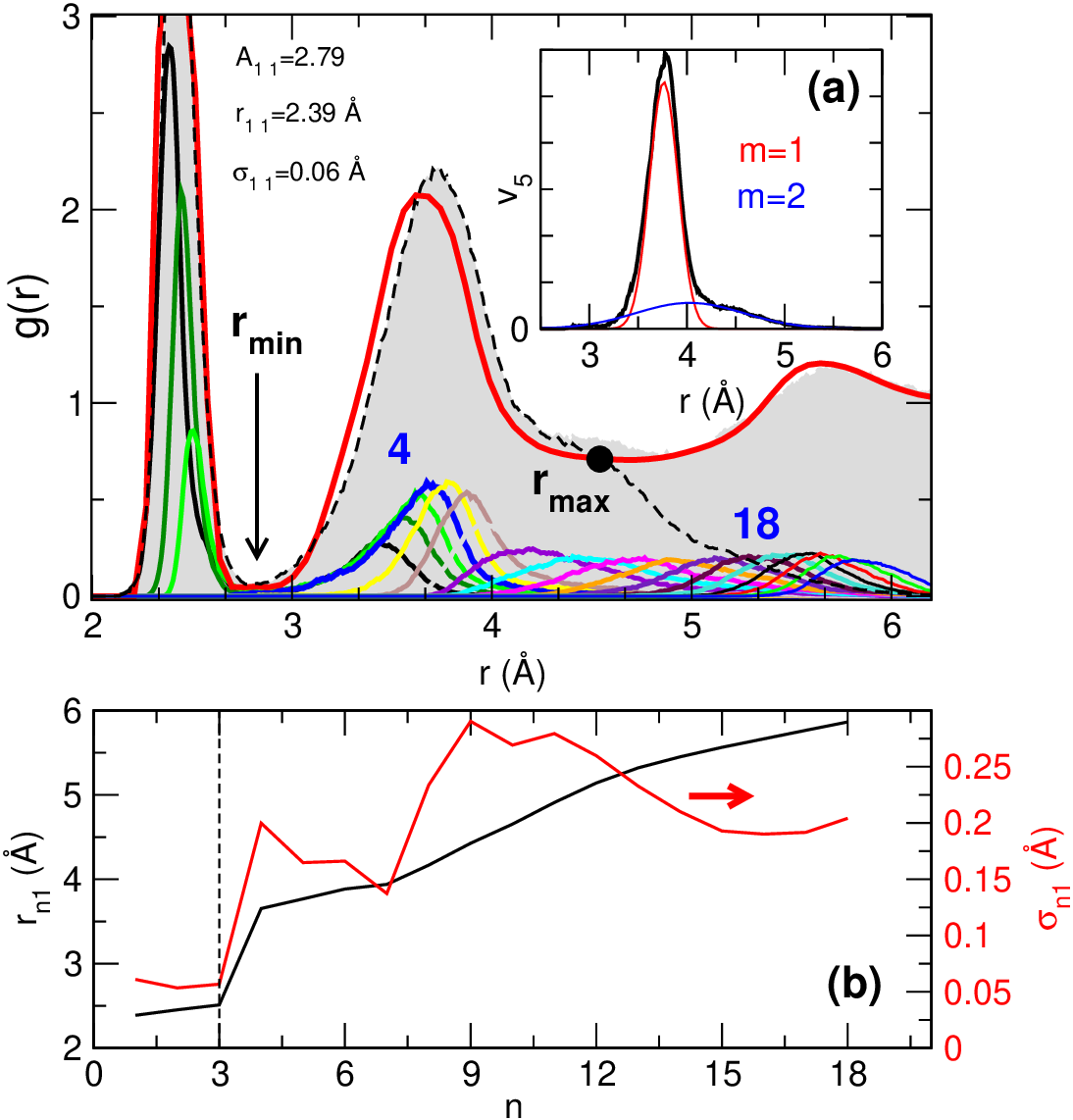}
\caption{\label{fig1} (Color online) a) Calculated total pair correlation function $g(r)$ (gray) of simulated amorphous As$_2$Se$_3$ \cite{asse} together with its decomposition into neighbor distribution functions $v_n(r)$ (colored curves, up to $n$=18) and corresponding results from neutron diffraction (red curve, \cite{salmon1}). The broken black curve corresponds to $\sum_{n=1}^{10}v_n(r)$ and defines a distance $r_{max}$ for a given $N$ (here $N$=10). The inset shows $v_5(r)$ and its decomposition into $M$=2 Gaussians. The distance $r_{min}$ defines the minimum of $g(r)$ and is used for the determination of a coordination number. b) Some parameters characterizing the first Gaussian decomposition $m=1$: mean distance $r_{n1}$ and variance $\sigma_{n1}$ for the main Gaussian decomposition ($m$=1). The broken line corresponds to $n$=3 and is linked with the dominant coordination number (As).}  
\end{figure}

We propose that the pair distribution function $g(r)$ can be decomposed into a series of neighbor distribution functions $v_n(r)$ (1$\leq n\leq$ N) as exemplified in different liquids and glasses \cite{neigh1,bychkov,neigh3} (Fig. \ref {fig1}a). For each atom, distributions are constructed from molecular dynamics (MD) based trajectories by sorting the neighbours according to the bond length. We assume that such distribution functions can be reconstructed by Gaussians $u_{nm}(r)$ with amplitudes $A_{nm}$, mean distances (first moments) $r_{nm}$ and variances $\sigma_{nm}^2$. Note that the index ($M\geq m\geq$1) signals that up to $M$ Gaussian functions might be used to fit a given distribution function $v_n(r)=\sum_m^Mu_{nm}(r)$ so that the total number of Gaussians is $N$+$M$-1. In the forthcoming, applications to different systems have shown that not more than $M$=3 Gaussians are needed to reproduce accurately a function $v_n(r)$ calculated from the MD simulations, and in most of the situations $M=1$ is sufficient, especially at short distance (see below).
\par
Using such distributions, a direct calculation shows that equ. (\ref{sk}) can be exactly computed and leads to the interference function:
\begin{eqnarray}
\label{central}
I(k)&=&k[S(k)-1]=\sum_{n=1}^NI_n(k)\\ \nonumber
&=&4\pi\rho_0\sum_{n,m}^{N,M}A_{nm}\sigma_{nm}^2\sqrt{2}e^{-r_{nm}^2/2\sigma_{nm}^2}\biggl[r_{nm}L_{nm}+k\sigma_{nm}^2V_{nm}\biggr]
\end{eqnarray}
where $V_{nm}$ and $L_{nm}$ represent the Voigt functions \cite{voigt} for the $m^{th}$ Gaussian component associated to the neighbour distribution $v_n(r)$ :
\begin{eqnarray}
\label{L}
L_{nm}&=&L_{nm}\biggl({\frac {k\sigma_{nm}}{\sqrt{2}}},-{\frac {r_{nm}}{\sigma_{nm}\sqrt{2}}}\biggr)\\
&=&{\frac {1}{\sqrt\pi}}\int_0^\infty
e^{-{t^2}/4+r_{nm}t/\sigma_{nm}\sqrt{2}}\sin\biggl({\frac {k\sigma_{nm}}{\sqrt{2}}}t\biggr)dt\nonumber
\end{eqnarray}
\begin{eqnarray}
V_{nm}&=&V_{nm}\biggl({\frac {k\sigma_{nm}}{\sqrt{2}}},-{\frac {r_{nm}}{\sigma_{nm}\sqrt{2}}} \biggr)\\
&=&{\frac {1}{\sqrt\pi}}\int_0^\infty
e^{-{t^2}/4+r_{nm}t/\sigma_{nm}\sqrt{2}}\cos\biggl({\frac {k\sigma_{nm}}{\sqrt{2}}}t\biggr)dt\nonumber
\end{eqnarray}
$V_{nm}$ and L$_{nm}$ being also equal to the real and imaginary parts of the Fadeeva function $w_n(z)$ \cite{expansion} with $z={\frac {k\sigma_{nm}}{\sqrt{2}}}-{\frac {ir_{nm}}{\sigma_{nm}\sqrt{2}}}$ defined from the complex error function :
\begin{eqnarray}
w_n(z)=e^{-z^2}\biggl[1+{\frac {2i}{\sqrt{\pi}}}\int_0^ze^{t^2}dt\biggr]
\end{eqnarray} 
In the thermodynamic limit, equation (\ref{central}) is an exact result and its validity can be checked for a variety of disordered systems with increasing $N$. 
\par
\begin{figure}
\includegraphics*[width=\linewidth, keepaspectratio=true]{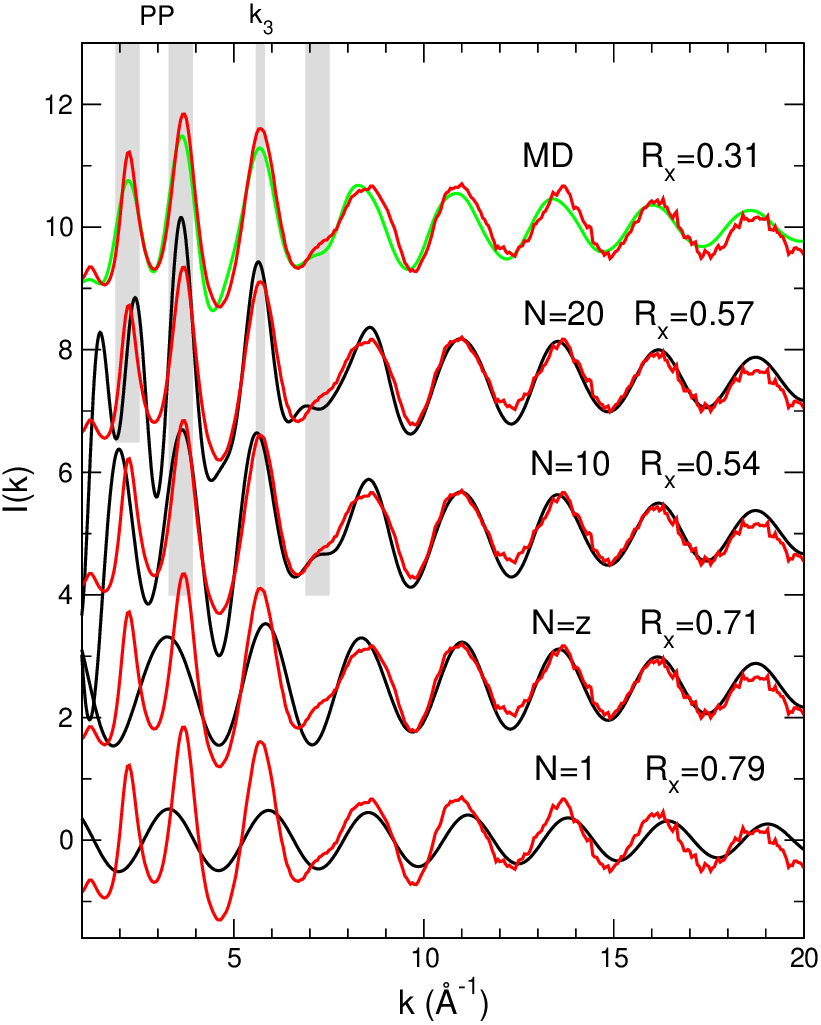}
\caption{\label{fig2a} (Color online) Reconstructed interference function $I(k)=~k[S(k)-1]$ (black curves) for increasing number $N$ of neighbor distributions in amorphous As$_2$Se$_3$, using equ. (\ref{central}) (duplicated and shifted red curves, exp. data \cite{salmon1}). The green curve corresponds to a calculated structure factor from MD simulations \cite{asse} using equ. (\ref{sk}). For As$_2$Se$_3$, one has $z$=3 neighbors around a central As atom. Gray zones associated with the main peaks are discussed in the text. The numbers $R_x$ represent the value of the Wright parameter which measures the degree of accuracy of equ. (\ref{central}).}  
\end{figure}
\par
First, we examine the convergence of the series expansion (equ. (\ref{central})) on amorphous As$_2$Se$_3$ but verify before that the pair correlation function $g(r)$ can be decomposed into such Gaussian neighbour functions $v_n$. Figure \ref{fig1}a shows a calculated pair correlation function obtained from First Principles MD simulations (top) \cite{asse} which reproduces very accurately experimental results from neutron scattering in real space \cite{salmon1}. The analysis up to 18 neighbors shows that corresponding neighbor functions $v_n(r)$ can be decomposed into Gaussian distributions (inset) with both $r_{nm}$ and $\sigma_{nm}$ increasing with $n$ (Fig. \ref{fig1}b), and a sharp increase for both quantities at $n=3$ signals the 3-fold coordinated As based network structure \cite{asse}. Such parameters are then used to recalculate the total structure factor using equ. (\ref{central}). Note that because there is an overlap between the different functions $v_n(r)$, at fixed $N$ the decomposition of $g(r)$ is only valid up to $r=r_{max}(N)$ (broken curve Fig. \ref{fig1}a). For a fixed $N$ and $r>r_{max}(N)$, additional neighbors are indeed missing (N+1,N+2,...) in order to have exactly $g(r)=\sum_nv_n(r)$, and this defines a minimal momentum transfer $k_{min}=2\pi/r_{max}$ for the validity of equ. (\ref{central}).
\par
Figure \ref{fig2a} shows the obtained results with increasing $N$ for the interference function $I(k)$. The latter permits to blow up the oscillations at large momentum transfer in order to check for the accuracy of the series expansion (\ref{central}). Quite obviously, a single neighbor decomposition $N$=1 (using the parameters given in Fig. \ref{fig1}b) leads only to a poor reproduction of $I(k)$, including the oscillations at large $k$. Instead, equ. (\ref{central}) reveals that such oscillations can be only obtained i) for multiple Gaussians having slightly different mean distances $r_{nm}$ and ii) when the number of Gaussians satisfies $N$+$M$-1=$z$, $z$ being related to the coordination number of the system and determined from the number of functions $u_{nm}(r)$ found for $r<r_{min}$ with $r_{min}$ the minimum of $g(r)$ (three in Fig. \ref{fig1}a). In the present case, a calculation of the coordination number $\bar n$ from such functions using the condition $N$+$M$-1$\leq z$ leads to $\bar n$=2.33 which is compatible with experiments ($\bar n$=2.30 \cite{salmon1}).
One also notices that for $k>$7~\AA$^{-1}$ any further evolution with increasing $N$ is barely visible up to $N$=20 (Fig. \ref{fig2a}), the contributions $I_n(k)$ with larger $n$ in this range of momentum transfer leading to $I_n(k)\simeq$0 \cite{suppl}. This results from the mathematical properties of the Fadeeva function which lead to i) $L_{nm}\gg V_{nm}$ for the considered parameters and range of momentum transfer $k$, and ii) to a rapid decay to zero of $L_{nm}$ with $k$, this effect being even enhanced with increasing $\sigma_{nm}$ due to the jump observed between first and second shell of neighbors (e.g. $n$=3 in Figure \ref{fig1}b). As a consequence, we arrive to the mathematical conclusion that the large $k$ behaviour of $S(k)$ is dominated by the first coordination shell or SRO of the system. 
\par
Using the series expansion, it is now insightful to analyze the experimental diffraction pattern \cite{salmon1} with growing $N$ in order to extract information on SRO and IRO elements. We find that the PP region (2-5~\AA$^{-1}$) depends on $N$, a near convergence being achieved for $N\simeq$20 and $I(k)$ has now a profile that is similar to the one calculated from MD \cite{asse}, albeit the secondary PP at 3.5~\AA$^{-1}$ is already correctly reproduced (position, amplitude) for $N$=10. The third prominent peak at $k_3$=5.8~\AA$^{-1}$ is obviously related to first and second neighbor shell atoms ($N\leq 10$) given the negligible contributions to $I_n(k_3)$ for 10$<n<$20 \cite{suppl}. The decomposition also reveals that a typical shoulder peak observed at 7.2~\AA$^{-1}$ can be unambiguously assigned with second shell correlations because it is absent for $N<z$, has a significant contribution for 4$<n<$7 and becomes negligibly small for larger $n$ \cite{suppl}. A standard means quantifying more precisely the accuracy of the approach builds on the Wright parameter $R_x$ which evaluates a squared deviation between experimental and theoretical data \cite{wright}. Corresponding numbers are given in Fig. \ref{fig2a}. Here, the increase of $N$ leads to a decrease of $R_x$ (albeit almost constant between $N$=10 and 20), thus indicating that the agreement of equ. (\ref{central}) should continuously improve with growing $N$. As a final comment, one should note that $r_{max}=$5.64~\AA\ for $N$=20 (and $k_{min}$= 1.1~\AA$^{-1}$). The reproduction of the FSDP region ($\simeq$1.5~\AA$^{-1}$) might be investigated but important oscillations remain for each contributions $n$ ($n\leq$20) so that a near convergence of equ. (\ref{central}) is not achieved for this range of momentum transfer. 
\par
\begin{figure}
\includegraphics*[width=\linewidth, keepaspectratio=true]{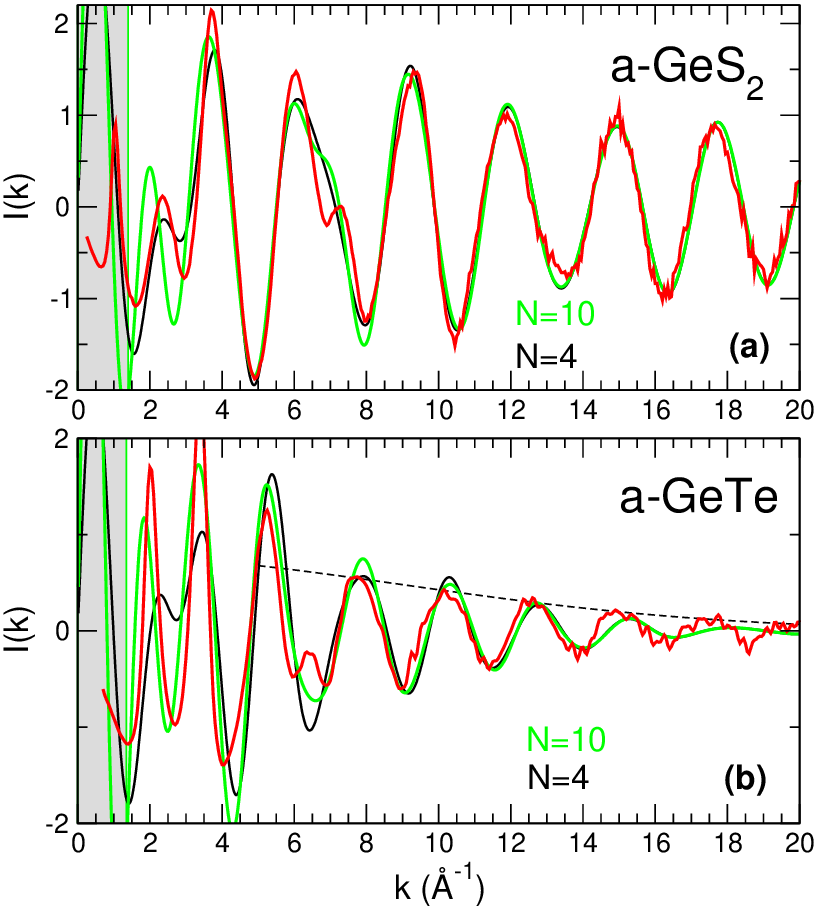}
\caption{\label{fig2} (Color online) a) Reconstructed interference functions $I(k)$ (red curves, exp. data) in amorphous a) GeS$_2$ (neutrons, \cite{salmon2}), b) GeTe (X-rays \cite{ghezzi}) with variable $N$ (black and green curves). The gray zone corresponds to $k<k_{min}=2\pi/r_{max}$. The broken line is an exponential decay $\exp[-k^2\sigma^2/2]$ with $\sigma$=0.12 (see text).}  
\end{figure}
The degree of generality of equ. (\ref{central}) can now be verified for other systems such as amorphous GeS$_2$ and GeTe. Figure \ref{fig2} shows the obtained results for the interference function $I(k)$ for two expansions ($N$=4 and 10). A close inspection reveals that the series expansion (equ. (\ref{central})) is largely material dependant because a large $k$ domain can be described only from the first neighbors ($N$=4) in GeS$_2$ whereas this is not the case for GeTe. Note that the quality of the MD simulations is nearly the same, as detected from the reproduction of the pair correlation function \cite{suppl}. However, the slight shift between theory and experiments occuring at large $k$ in a-GeTe is indicative of the well-known bond length mismatch as theory usually overestimates the Ge-Te bond distance \cite{gete_vdw} with respect to experiments \cite{ghezzi}. For this system, the two peaks of the PP region (1.8-4~\AA$^{-1}$) is conveniently described by second shell neighbor distributions ($N$=10).
\par
An exact expression for the high momentum transfer limit can be obtained by performing an expansion \cite{suppl} of the Fadeeva function \cite{expansion} (equ. (\ref{central})) at $k\rightarrow\infty$, and $I_n(k)$ behaves as :
\begin{eqnarray}
I_n(k)\simeq 8\pi\rho_0\sqrt{2}\sum_mA_{nm}r_{nm}\sigma_{nm}e^{-k^2\sigma_{nm}^2/2}\sin{kr_{nm}}
\end{eqnarray}
where the terms beyond the first shell are negligible because of the jump increase of $\sigma_{nm}$, and furthermore $m=1$ (single Gaussians for a given $n$). If one also assumes that $\sigma_{n1}$ is nearly constant for the first shell of neighbors ($\sigma_{n1}\simeq\sigma$, Fig. \ref{fig1}b) and $A_{m1}\simeq A$ by convenience, one obtains the compact form :
\begin{eqnarray}
\label{approx}
I(k)\simeq 8\sqrt{2}\pi\rho_0A\sigma e^{-k^2\sigma^2/2}{\frac {d}{dk}}\ \ {\frac {\cos{k\bar r}\sin{kza}}{\sin{{\frac {ka}{2}}}}}
\end{eqnarray}
\begin{table}[b]
\caption{Calculated average distance $\bar r$ using the explicit perioditicity $\Lambda$ of the large $k$ behavior of $I(k)$ (see text), determined for different amorphous and liquid materials. It is compared to the bond distance $d$ determined from the principal peak position of the experimental pair correlation function $g(r)$.}
\vspace{0.3cm}
\begin{ruledtabular}
\begin{tabular}{lccc}
System&$\bar r$ (\AA)& $d$ (\AA)&Reference \\
\colrule
\\
a-As$_2$Se$_3$&2.37$\pm$0.17&2.39&\cite{salmon1} \\
a-GeS$_2$&2.22$\pm$0.16&2.21&\cite{salmon2} \\
a-GeSe$_2$&2.32$\pm$0.04&2.36&\cite{Nature_salmon} \\
%a-Si$O_2$&1.97&3.36&\cite{xx} \\
a-GeTe&2.66$\pm$0.07&2.61&\cite{ghezzi} \\
l-GeTe$_4$ (820 K)&2.68$\pm$0.04&2.73&\cite{prb2014} \\
l-H$_2$O (300 K)&2.80$\pm$0.15&2.81&\cite{sope}  \\
l-Te (623 K)&2.68$\pm$0.05&2.77&\cite{bychkov}
\\ \colrule
\\
\end{tabular}
\end{ruledtabular}
\label{table2}
\end{table}

where one has assumed that $r_{n1}$ increases linearly in the first shell ($r_{n1}=r_0+na$, Fig. \ref{fig1}b) where $a$ is a bond distance increment (typically $a$=0.06~\AA) and $\bar r=z^{-1}\sum_{n=1}^zr_{n1}$ is the average bond distance. Given the value of the different atomic parameters, the arguments of the trigonometric functions have the property $za\ll \bar r$ so that the periodicity $\Lambda$ of the interference function at large $k$ is dominated by $\Lambda$=$2\pi/\bar r$ and connects to the SRO characteristics (average first neighbor distance) of the material. Corresponding results determined from experimental data are given in Table \ref{table2} and provide a good agreement with the distances $d$ that are directly obtained from the principal peak position of the measured pair correlation function $g(r)$. The modulation in equ. (\ref{approx}) has a periodicity of $\Lambda_m$=$\pi/2za\gg\Lambda$ which provides a measure of both the coordination number and $a$ but such a modulation is barely visible from the considered $k$ range (Fig. \ref{fig2}). The exponential decay $\exp{-k^2\sigma^2/2}$ provides information on the spatial extent of the neighbor distributions and also a measure of the rigidity of the SRO geometrical units because a small $\sigma$ value will induce a small bond variability. Qualitatively, the decay of $I(k)$ for amorphous GeTe (Fig. \ref{fig2}) can now be linked with the presence of more softer units which leads to larger values for $\sigma$ as also independently verified from the calculated Ge-centred bond angle distribution which exhibits a broader distribution around 109$^\circ$, as compared to selenides or sulphides \cite{gete_vdw}. Using equ. (\ref{approx}), a fit to the data at large $k$ at the maximum of the oscillations up to the maximal available $k$-range ($\simeq$20-35~\AA$^{-1}$) \cite{salmon1,salmon2,ghezzi} leads to $\sigma$=0.098~\AA, 0.063~\AA, 0.12~\AA\ for As$_2$Se$_3$, GeS$_2$, GeTe respectively, smaller than the one performed on the experimental data of water ($\sigma$=0.163~\AA, \cite{sope}). An increased exponential decay of $I(k)$ (equ. (\ref{approx})) is, indeed, visible in the liquid state \cite{bychkov,prb2014,sope}, and is driven by the larger value of $\sigma$ induced by the increased atomic motion which broadens the typical peaks of the pair correlation function. 
\par
In the low $k$ limit, a similar expansion of equ. (\ref{central}) permits to access to the long wavelength limit ($k\simeq$0) and $S(k)$ that can be expanded from equs. (\ref{central}) and (\ref{L}).
\begin{eqnarray}
S(0)&\simeq& 1+{\frac {8\rho_0}{\sqrt{\pi}}}\sum_{n,m}^N A_{nm}\sigma_{nm}^3r_{nm}\biggl(e^{-{r_{nm}}^2/2\sigma_{nm}^2}\\ \nonumber
&+&{\frac {\sqrt{2\pi}}{\sigma_{nm}}}r_{nm}erfc\biggl[-{\frac {r_{nm}}{\sigma_{nm}\sqrt{2}}}\biggr]\biggr)
\end{eqnarray}
where, by virtue of the definition of $S(0)$ from number fluctuations \cite{hansen}, it is only valid in the thermodynamic limit ($N\rightarrow\infty$) \cite{thorpe,mousseau}. 
\par
In summary, we have shown that the numerical decomposition of the pair distribution function into neighbor distribution functions permits to reconstruct the structure factor $S(k)$ as a series expansion, in a fashion that bears similarities with the pioneering work on crystals with weak disorder \cite{warren,bernal,guinier}. Results not only indicate that a few shells of neighbors ($\simeq$~20 atoms) are sufficient in order to describe $S(k)$ over extended ranges in momentum transfer, they also provide a direct link between typical features observed in reciprocal space and the neighbour rank (i.e. $N$) which should certainly help for an improved analysis of experimental diffraction patterns of glasses. It would certainly be interesting to extend such an analysis to partial correlations and to other typical glass-forming materials or liquids. Work in this direction is in progress.  
\par
{\bf Acknowledgements :} The author thanks C. Benmore, E. Bychkov, L. Cormier, G.J. Cuello, P.S. Salmon and A. Zeidler for useful discussions and for sharing their experimental data. French HPC Center ROMEO of Universit\'e Reims Champagne-Ardenne is acknowledged for supercomputing access.

\end{document}